\documentclass[aps,preprint,nofootinbib,superscriptaddress]{revtex4-1}
\usepackage{amssymb}

\usepackage{epsfig}
\usepackage[bookmarksnumbered,bookmarksopen,colorlinks,citecolor=blue,linkcolor=blue]{hyperref}

\usepackage{amsmath}

\begin{document}


\title{Mass dependence of symmetry energy coefficients in Skyrme force}

\author{N. Wang}
\email{wangning@gxnu.edu.cn}\affiliation{ Department of Physics,
Guangxi Normal University, Guilin 541004, People's Republic of
China }
\affiliation{ State Key Laboratory of Theoretical Physics, Institute of Theoretical Physics, Chinese Academy of Sciences, Beijing 100190, China}

\author{M. Liu}
\affiliation{ Department of Physics,
Guangxi Normal University, Guilin 541004, People's Republic of
China }

\author{H. Jiang}
\affiliation{School of Arts and Sciences, Shanghai Maritime University, Shanghai 201306, People¡¯s Republic of China }

\author{J. L. Tian}
\affiliation{School of Physics and Electrical Engineering, Anyang Normal University, Anyang 455000, People¡¯s Republic of China}

\author{Y. M. Zhao}
\affiliation{Department of Physics and Astronomy, and Shanghai Key Laboratory for Particle 
Physics and Cosmology, Shanghai Jiao Tong University, Shanghai 200240, People¡¯s Republic of China}

\affiliation{ IFSA Collaborative Innovation Center, Shanghai Jiao Tong University, Shanghai 200240, People¡¯s Republic of China
 }

\begin{abstract}
Based on the semi-classical extended Thomas-Fermi approach, we study the mass dependence of the symmetry energy coefficients of finite nuclei for 36 different Skyrme forces. The reference densities of both light and heavy nuclei are obtained. Eight models based on nuclear liquid drop concept and the Skyrme force SkM* suggest the symmetry energy coefficient $a_{\rm sym}=22.90 \pm 0.15 $ MeV at $A=260$, and the corresponding reference density is $\rho_A\simeq 0.1$ fm$^{-3}$ at this mass region. The standard Skyrme energy density functionals give negative values for the coefficient of the $I^4$ term in the binding energy formula, whereas the latest Weizs\"acker-Skyrme formula and the experimental data suggest positive values for the coefficient.

\end{abstract}

\maketitle

\begin{center}
\textbf{I. INTRODUCTION}
\end{center}

Nuclear symmetry energy has attracted a lot attention in recent years. In addition to its importance in the study of nuclear physics such as nuclear exotic structures and reactions induced by unstable nuclei, the symmetry energy also plays a role in the study of nuclear astrophysics, such as the r-process and the properties of neutron stars \cite{Li14}. Although a great effort has been devoted in recent decades to investigate the symmetry energy \cite{Li08,Chen05,Shet07,Botvina02,Stein,Stein05,Dong11}, the density dependence of the symmetry energy, even at sub-saturation density region, is not very well constrained. One usually obtains the information of the symmetry energy from nuclear dynamical behavior in reactions \cite{Zhang08,Tsang09,Trip08} and the static properties of finite nuclei such as nuclear masses \cite{Wang10,Wang14,HFB17,HFB27,Zhao10,Liu2010,Jiang12,Tian14} and neutron skin thickness \cite{Cent09,Wang13,Radii,Chen13}, or from the asymmetric nuclear matter based on various effective interactions \cite{Zuo14,Vret03,Gre03}, or from the observables in nuclear astrophysics. To understand the behavior of symmetry energy in nuclear masses and nuclear matter, it is crucial to establish a reliable connection between the mass dependence of the symmetry energy coefficient $a_{\rm sym} (A)$ of finite nuclei and the density dependence of the symmetry energy $E_{\rm sym} (\rho)$ of nuclear matter, with which the obtained symmetry energy coefficients from various macroscopic-microscopic or liquid-drop models could be used to constrain the behavior of $E_{\rm sym} (\rho)$ at sub-saturation densities.

It is known that the density functional theory is widely used in the study of the nuclear ground state which provides us with a useful balance between accuracy and computation cost, allowing large systems with a simple self-consistent manner. With the same energy density functional, both the $E_{\rm sym} (\rho)$ and $a_{\rm sym} (A)$ could be self-consistently investigated. In the framework of the effective Skyrme interaction \cite{Vautherin} and the extended Thomas-Fermi (ETF) approximation, the Skyrme energy density functional approach was proposed for the study of nuclear structure and reactions \cite{Bencheikh, M.Brack,liumin}. One of the advantages in this approach is that the Coulomb energy, the Wigner energy and the microscopic shell corrections can be "cleanly" removed from the binding energies of nuclei, which is important to accurately obtain the information of symmetry energy. Another advantage of this approach is that the higher-order terms of the symmetry energy can be investigated simultaneously.

On the other hand, the investigation of symmetry energy is helpful to improve and test the reliability of various nuclear mass models, especially for the predictions of the masses of nuclei approaching neutron drip line. Both the macroscopic-microscopic mass formulas \cite{Wang10,Wang14} and the microscopic Hartree-Fock-Bogoliubov (HFB) models \cite{HFB17,HFB27} can reproduce the measured masses of more than 2000 nuclei with an rms error of several hundred keV. However, the predictions for extremely neutron-rich nuclei are quite different from these models although the similar symmetry energy coefficient is adopted. For example, adopting almost the same symmetry energy coefficient $J\approx 30$ MeV, the predicted mass of $^{168}$Sn with the WS4 model \cite{Wang14} is larger than those with the HFB17 model \cite{HFB17} by 24.4 MeV, which is difficult to be explained from the microscopic shell and pairing corrections. The higher-order term of symmetry energy might play a relevant role to the large difference of the predicted masses, in addition to other effects. It is therefore necessary to investigate the symmetry energy coefficient of finite nuclei especially the coefficient of the $I^4$ term in the standard Skyrme force.

\begin{center}
\textbf{II. THE THEORETICAL MODEL}
\end{center}

The nuclear energy part of a nucleus at its ground state can be expressed as the
integral of the Skyrme energy density functional $\mathcal{H}$({\bf r}):
\begin{eqnarray}
E = \int {\mathcal H} [\rho_{\rm q}({\bf r}), \tau_{\rm q}({\bf r}), {\bf J}_{\rm q}({\bf r})] \; d{\bf r},
\end{eqnarray}
with the local nucleon densities $\rho_{\rm q}({\bf r})$, kinetic energy densities $\tau_{\rm q}({\bf r})$ and spin-orbit densities ${\bf J}_{\rm q}({\bf r})$ ($q=n$ for neutrons and $q=p$ for protons). By using the extended Thomas-Fermi approach up to the second order in $\hbar$ (ETF2) \cite{Bencheikh, M.Brack,liumin}, the kinetic energy density and the spin-orbit density can be expressed as a functional of nuclear density and its gradients. In the semi-classical ETF2 approach, one can obtain the "macroscopic" part of the nuclear energy $\tilde{E}[\rho_n({\bf r}), \rho_p({\bf r})]$ of a nucleus. The microscopic shell, pairing, and Wigner effects are not involved in this semi-classical approach, which is helpful to study the symmetry energy coefficients of finite nuclei as cleanly as possible.

To remove the influence of nuclear deformations, we use the spherical Fermi function
\begin{eqnarray}
\rho_q ({\bf r} )= \frac{\rho^{(q)}_{0}}{1+  \exp (\frac{  {\bf r} -R_{q} }{a_q})},
\end{eqnarray}
for describing the density of a nucleus. $\rho^{(q)}_{0}$, $R_q$ and $a_{q}$ denote the central density, radius, and surface diffuseness of nuclei, respectively. The central density is determined from the conservation of particle number. By using the optimization algorithm \cite{liumin} and varying the four variables $R_{p}$, $a_{p}$, $R_{n}$, $a_{n}$ in Eq.(2) for a given nucleus, one can self-consistently obtain the minimal "macroscopic" energy $\tilde{E}$ of the nucleus. It is found that the ETF approach can reproduce quite accurately the Hartree-Fock average field and corresponding density distribution of $^{208}$Pb \cite{Bencheikh}. As the same as those did in Ref. \cite{Rein06}, the calculations have been carried out for nuclei with huge numbers of nucleons, of the order of $10^6$, in order to perform a reliable extrapolation in the inverse radius, and the Coulomb interaction has been ignored to be able to approach nuclei of arbitrary sizes and to avoid radial instabilities characteristic of systems with very large atomic numbers.

\begin{center}
\textbf{III. RESULTS AND DISCUSSIONS}
\end{center}

\begin{figure}
\includegraphics[angle=-0,width= 0.7 \textwidth]{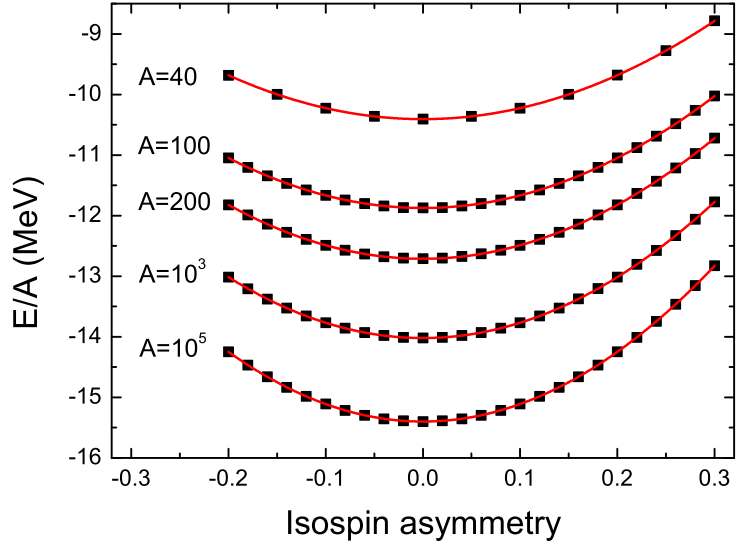}
 \caption{(Color online) Binding energy per particle as a function of $I=(N-Z)/A$. The Coulomb interaction is not involved in all calculations with the ETF2 approach. The squares denote the results with the Skyrme force SkM*, and the curves denote the fit with Eq.(3).}
\end{figure}

As an example, "macroscopic" energy per particle of nuclei is studied by adopting the Skyrme force SkM* \cite{SkMs} together with the ETF2 approach, since SkM* is very successful for describing the bulk properties and surface properties of nuclei. Figure 1 shows the calculated "macroscopic" energy per particle $\tilde{E}/A$ as a function of isospin asymmetry $I=(N-Z)/A$ for nuclei with given mass number $A$.   The calculations are performed in the region of $-0.2\le I \le 0.3$ which generally covers all nuclei with known masses. The squares denote the calculated results with the Skyrme force SkM*. Ignoring the Coulomb energy and microscopic corrections, the binding energy per particle of nuclei is usually written as
\begin{eqnarray}
\tilde{E}/A=e_0(A)+a_{\rm sym}(A) I^2 + a_{\rm sym}^{(4)}(A) I^4+ \mathcal{O}(I^6) .
\end{eqnarray}
The solid curves in Fig. 1 denote the fit to the squares by using the form of Eq.(3). The results of SkM* can be remarkably well reproduced by the solid curves, with the rms deviations smaller than $10^{-6}$ MeV. Figure 2 shows the obtained symmetry energy coefficients of finite nuclei as a function of mass number. The squares and circles denote the extracted results from the Skyrme force SkM* for the values of $a_{\rm sym}$ and $a_{\rm sym}^{(4)}$, respectively. We find that the values of $a_{\rm sym}^{(4)}$ are negative from the calculations of SkM* and the values of $a_{\rm sym}$ approach 30 MeV with increasing of mass number. To check the convergence of Eq.(3), we re-calculate the values of $a_{\rm sym}$ by neglecting the $I^4$ and higher-order terms in Eq.(3) and find that the obtained values of $a_{\rm sym}$ are slightly reduced by about 0.2 MeV.

\begin{figure}
\includegraphics[angle=-0,width= 0.7 \textwidth]{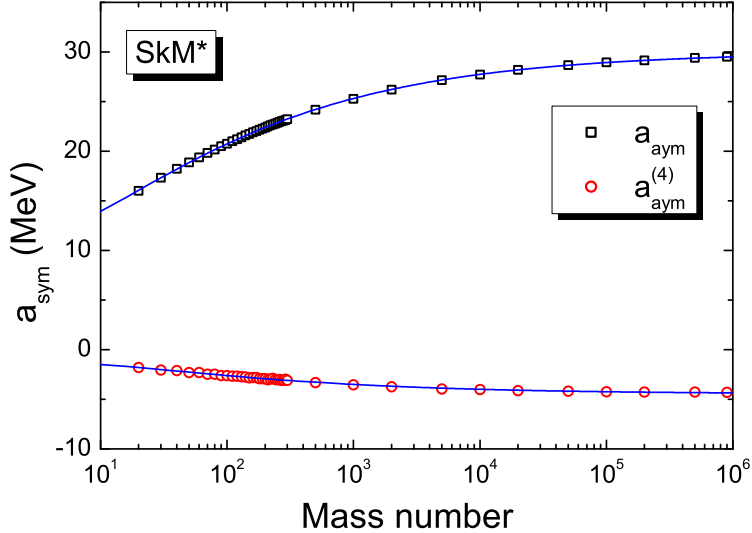}
 \caption{(Color online) Symmetry energy coefficients as a function of mass number of nuclei.}
\end{figure}

To further check the asymptotic behavior of the symmetry energy coefficient when $A\to \infty$, we further use a Polynomial fit to reproduce the calculated $a_{\rm sym}$ and $a_{\rm sym}^{(4)}$. The solid curves in Fig. 2 denote the results of the Polynomial fit to the scattered symbols. To see the results more clearly, the results for $a_{\rm sym}$ with SkM* (open squares) and with SLy4 \cite{SLy4567} (open circles) are shown in Fig. 3 as a function of $A^{-1/3}$. According to the liquid drop model, the symmetry energy coefficient of a finite nucleus is usually written as
\begin{eqnarray}
a_{\rm sym}(A)=J-a_{\rm ss}A^{-1/3}+a_{\rm cs} A^{-2/3}
\end{eqnarray}
by using the Leptodermous expansion in terms of powers of $A^{-1/3}$. $J\approx 28 -34$ MeV denotes the symmetry energy of nuclear matter at normal density. $a_{\rm ss}$ is the coefficient of the surface-symmetry term and $a_{\rm cs}$ denotes the curvature-symmetry term. Through the Polynomial fit to the results of different Skyrme forces, one can obtain the corresponding values of $J$, $a_{\rm ss}$ and $a_{\rm cs}$.

\begin{figure}
\includegraphics[angle=-0,width= 0.7 \textwidth]{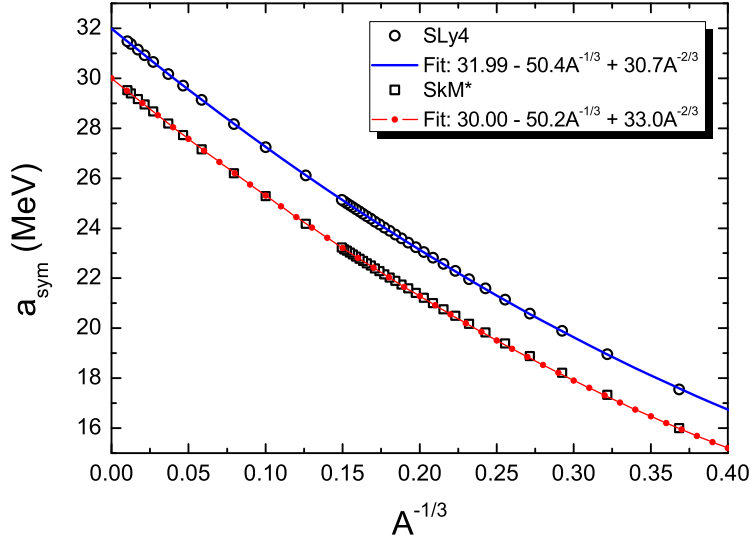}
 \caption{(Color online) Symmetry energy coefficient $a_{\rm sym}$ as a function of $A^{-1/3}$. The open circles and squares denote the results of ETF2 plus SLy4 and SkM*, respectively. The curves denote the Fit with Eq.(4).}
\end{figure}

For nuclear matter, the symmetry energy in the standard Skyrme energy density functional is expressed as
\begin{eqnarray}
E_{\rm sym}(\rho) &=& \frac{1}{3}\frac{\hbar^2}{2m} \left ( \frac{3\pi^2}{2}
\right )^{2/3} \rho^{2/3}-  \frac{1}{8} t_0 (2 x_0+1)\rho  \nonumber \\
&-&   \frac{1}{24} \left(
 \frac{3 \pi^2}{2} \right )^{2/3} \Theta_{\rm sym}
 \rho^{5/3}-\frac{1}{48} t_3 (2x_3+1) \rho^{\sigma+1}
\end{eqnarray}
with $ \Theta_{\rm sym}=3 t_1 x_1 -t_2(4+5 x_2)$. $t_0$, $t_1$,
$t_2$, $t_3$, $x_0$, $x_1$, $x_2$, $x_3$ and $\sigma$  are the Skyrme parameters. For the parameter set SkM* and SLy4, the corresponding symmetry energy $E_{\rm sym} (\rho_0)$ at saturation density $\rho_0$ from Eq.(5) is 30.0 and 32.0 MeV, respectively. From Fig. 3, we also note that the obtained value of $J$ from the symmetry energy coefficient of finite nuclei in the ETF2 approach is very close to the corresponding value of $E_{\rm sym} (\rho_0)$. Furthermore, the obtained binding energies per particle $e_0 (\infty)$ from the extrapolation with the Polynomial fit, i.e., the lowest energy in Fig. 1 when the mass number $A \to \infty $, are $-15.78$ MeV for SkM* and $-15.99$ MeV for SLy4, which are also very close to the corresponding energy per nucleon $E/A(\rho_0)$ at saturation density of asymmetric nuclear matter \cite{Dut12}.

\begin{table}
\centering
\caption{Properties of nuclear symmetry energy with the ETF2 approach (in MeV). }
\begin{tabular}{lcccccccc}
 \hline\hline
Force & ~~~~$L_c$~~~~ & ~~~~ $J$ ~~~~ & ~~~~$a_{\rm ss}$~~~~ & ~~~~$a_{\rm cs}$~~~ & ~$a_{\rm sym}^{(4)}  (\infty)$ ~ & ~$a_{\rm sym}^{(4)}  (208)$ ~ & ~$a_{\rm sym}^{(4)}  (40)$ ~ &  \\
\hline
BSk1 \cite{BSk1}	&	30.31 	&	27.95 	&	28.9 	&	7.3 	&	0.3 	&	-0.8 	&	-1.4 	     \\
MSk2 \cite{MSk2}	&	39.66 	&	29.99 	&	40.0 	&	19.2 	&	-2.2 	&	-2.6 	&	-2.6 	 	 \\
MSk6 \cite{MSk2}	&	30.92 	&	28.01 	&	29.3 	&	7.8 	&	0.2 	&	-0.8 	&	-1.4 	 	 \\
RATP \cite{RATP}	&	38.06 	&	29.25 	&	40.5 	&	20.2 	&	-1.4 	&	-2.3 	&	-2.4 		 \\
SkM	 \cite{SkM}     &	44.55 	&	30.71 	&	50.1 	&	32.9 	&	-5.2 	&	-3.5 	&	-2.6 	 	 \\
SkM* \cite{SkMs}	&	42.93 	&	30.00 	&	50.2 	&	33.0 	&	-4.4 	&	-2.9 	&	-2.2 		 \\
SkMP \cite{SkMP}	&	49.64 	&	29.79 	&	53.3 	&	40.2 	&	-10.0 	&	-5.2 	&	-3.3 	 	 \\
SkSC1 \cite{SkSC1}	&	28.87 	&	28.11 	&	26.7 	&	4.6 	&	0.5 	&	-0.4 	&	-1.2 		 \\
SkSC4 \cite{SkSC4}	&	29.06 	&	28.81 	&	27.0 	&	4.0 	&	0.5 	&	-0.5 	&	-1.4 		 \\
SkT1* \cite{SkT}	&	47.69 	&	31.97 	&	56.0 	&	39.4 	&	-6.6 	&	-4.7 	&	-3.4 		 \\
SkT3 \cite{SkT}	    &	46.80 	&	31.46 	&	48.7 	&	29.7 	&	-6.4 	&	-5.4 	&	-4.4 	 	 \\
SkT3* \cite{SkT}	&	47.25 	&	31.64 	&	52.6 	&	35.0 	&	-6.6 	&	-5.4 	&	-4.3 		 \\
SkT4 \cite{SkT}	    &	61.79 	&	35.24 	&	78.1 	&	71.0 	&	-18.3 	&	-8.5 	&	-4.8 	 	 \\
SkT6 \cite{SkT}	    &	38.85 	&	29.96 	&	41.2 	&	20.1 	&	-1.8 	&	-2.4 	&	-2.5 		 \\
SkT7 \cite{SkT}	    &	38.68 	&	29.51 	&	40.1 	&	19.7 	&	-1.5 	&	-2.4 	&	-2.6 		 \\
SkT8 \cite{SkT}	    &	38.81 	&	29.92 	&	41.4 	&	20.8 	&	-2.1 	&	-2.5 	&	-2.5 		 \\
Skz2 \cite{Skz} 	&	35.24 	&	32.02 	&	37.5 	&	13.6 	&	0.2 	&	-1.2 	&	-1.9 	 	 \\
Skz3 \cite{Skz} 	&	32.22 	&	32.03 	&	31.8 	&	6.6 	&	0.1 	&	-1.6 	&	-2.6 	 	 \\
Skz4 \cite{Skz} 	&	28.51 	&	32.02 	&	26.0 	&	0.3 	&	0.1 	&	-1.7 	&	-3.0 	 	 \\
SLy0 \cite{SLy012389}	&	41.90 	&	31.97 	&	49.3 	&	29.1 	&	-4.7 	&	-3.5 	&	-2.8 		 \\
SLy1 \cite{SLy012389}	&	41.89 	&	31.97 	&	51.4 	&	32.0 	&	-4.7 	&	-3.0 	&	-2.2 		 \\
SLy2 \cite{SLy012389}	&	41.99 	&	31.99 	&	47.4 	&	26.5 	&	-4.7 	&	-4.1 	&	-3.5 		 \\
SLy3 \cite{SLy012389}	&	41.34 	&	31.98 	&	50.0 	&	30.1 	&	-4.3 	&	-3.0 	&	-2.3 		 \\
SLy4 \cite{SLy4567}	&	41.59 	&	31.99 	&	50.4 	&	30.7 	&	-4.5 	&	-3.0 	&	-2.3 		 \\
SLy5 \cite{SLy4567}	&	42.19 	&	31.98 	&	51.6 	&	32.2 	&	-4.9 	&	-3.1 	&	-2.2 		 \\
SLy6 \cite{SLy4567}	&	41.96 	&	31.94 	&	47.6 	&	27.6 	&	-4.8 	&	-3.4 	&	-2.7 	 	 \\
SLy7 \cite{SLy4567}	&	41.86 	&	31.98 	&	46.6 	&	26.2 	&	-4.7 	&	-3.5 	&	-2.8 	 	 \\

\hline
\end{tabular}
\end{table}

\begin{table}
\centering
\caption{(\emph{Continued.}) }
\begin{tabular}{lcccccccc}
 \hline\hline
Force & ~~~~$L_c$~~~~ & ~~~~ $J$ ~~~~ & ~~~~$a_{\rm ss}$~~~~ & ~~~~$a_{\rm cs}$~~~ & ~$a_{\rm sym}^{(4)}  (\infty)$ ~ & ~$a_{\rm sym}^{(4)}  (208)$ ~ & ~$a_{\rm sym}^{(4)}  (40)$ ~ &  \\
\hline
SLy8 \cite{SLy012389}	&	41.89 	&	31.98 	&	50.6 	&	31.1 	&	-4.7 	&	-3.1 	&	-2.3 	 \\
SLy9 \cite{SLy012389}	&	44.95 	&	31.95 	&	51.0 	&	32.7 	&	-6.4 	&	-4.3 	&	-3.2 	 \\
SLy10 \cite{SLy4567}	&	39.39 	&	31.98 	&	41.4 	&	19.7 	&	-3.1 	&	-3.4 	&	-3.3 	 \\
SLy230a	\cite{SLy230} &	39.39 	&	31.96 	&	45.2 	&	23.4 	&	-4.4 	&	-3.5 	&	-3.0 		 \\
SLy230b	\cite{SLy230} &	41.60 	&	31.99 	&	50.4 	&	30.7 	&	-4.4 	&	-3.0 	&	-2.3 		 \\
SV-sym32 \cite{SVsym32} &	49.29 	&	32.12 	&	50.5 	&	32.5 	&	-6.7 	&	-6.0 	&	-5.0 	 \\
V080 \cite{V080}	&	30.12 	&	28.01 	&	29.8 	&	8.7 	&	0.8 	&	-0.2 	&	-0.9 		 \\
V090 \cite{V080}	&	30.31 	&	28.01 	&	29.0 	&	7.6 	&	0.6 	&	-0.5 	&	-1.2 		 \\
V110 \cite{V080}	&	30.15 	&	28.01 	&	28.7 	&	7.1 	&	0.3 	&	-0.7 	&	-1.4 		 \\
\hline
\end{tabular}
\end{table}

In this work, we have performed a systematical study of the symmetry energy coefficient for 36 different Skyrme forces in which the corresponding incompressibility coefficient for symmetry nuclear matter is about $K_\infty=240 \pm 30$ MeV and the measured masses of 54 spherical nuclei according to the predicted shapes of nuclei from the WS4 model can be roughly reproduced by using the ETF2 approach (with the rms deviations to the masses of the 54 nuclei less than 10 MeV). The results are listed in Table  I. We find that the average deviation between $J$ obtained from the symmetry energy coefficient of finite nuclei and $E_{\rm sym} (\rho_0)$ from Eq.(5) is only 0.04 MeV. In addition, it is important and necessary to check the obtained coefficient $a_{\rm sym}^{(4)}$ of the $I^4$ term from the symmetry energy coefficients of finite nuclei. The fourth-order term of the energy per nucleon at saturation density $E_{\rm sat,4}$ of asymmetric nuclear matter were systematically investigated by Lie-Wen Chen et al. for different Skyrme forces \cite{Chen09}. With the Polynomial fit to the calculated $a_{\rm sym}^{(4)}$ for a certain Skyrme force, one can obtain the corresponding value of $a_{\rm sym}^{(4)}(\infty)$ which is also listed in Table I. We find that $a_{\rm sym}^{(4)}(\infty)$ obtained in the ETF2 approach is close to the corresponding value of $E_{\rm sat,4}$. The average deviation between $a_{\rm sym}^{(4)}(\infty)$ extracted in this work and $E_{\rm sat,4}$ given in Ref. \cite{Chen09} is only 0.23 MeV. These tests for $J$,  $e_0 (\infty)$ and $a_{\rm sym}^{(4)}(\infty)$ indicate that the ETF2 approach is reliable for the study of the symmetry energy coefficient of finite nuclei.

\begin{figure}
\includegraphics[angle=-0,width= 0.7 \textwidth]{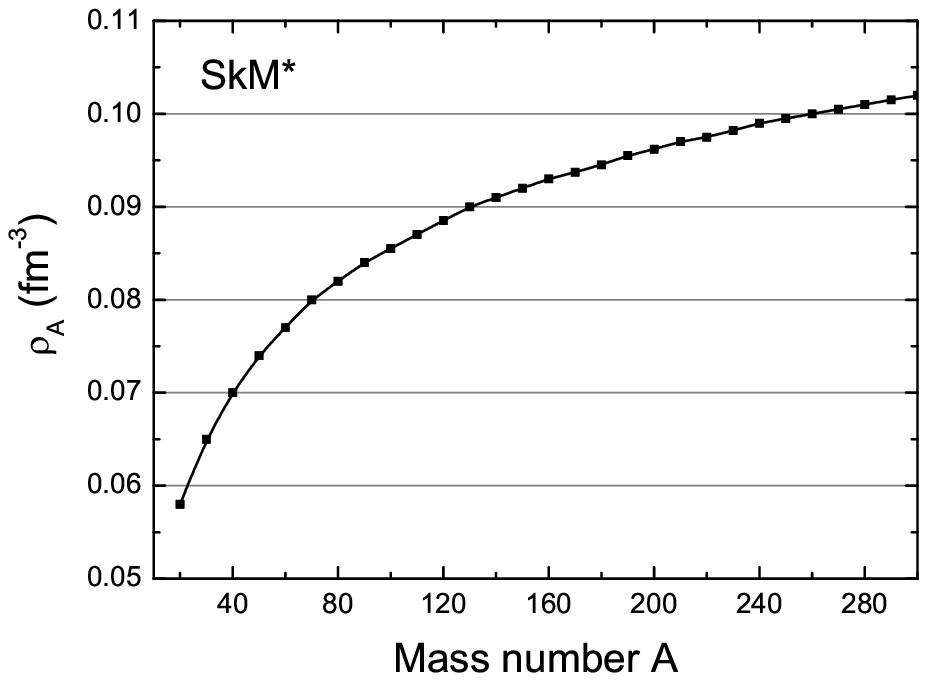}
 \caption{ Reference density obtained from SkM* as a function of mass number of nuclei.}
\end{figure}

With the ETF2 approach, the connection between the symmetry energy coefficient $a_{\rm sym}(A)$ and the symmetry energy of nuclear matter $E_{\rm sym}(\rho)$ can be established by using the relationship $a_{\rm sym}(A)= E_{\rm sym}(\rho_A)$ \cite{Cent09}, where $\rho_A$ is the reference density. It is usually thought that the reference density of a heavy nucleus such as $^{208}$Pb is about 0.1 fm$^{-3}$ \cite{Liu2010,Cent09}. In Fig. 4, we show the calculated reference density $\rho_A$ with the ETF2 approach by adopting the parameter set SkM*. According to the results of SkM*, we note that the reference density for nuclei with $A=260$ is about 0.1 fm$^{-3}$. In Table I, we also list the slope $L_c= 3\rho_c \left (\frac{\partial E_{\rm sym}}{\partial \rho} \right)_{\rho=\rho_c}$ of the symmetry energy $E_{\rm sym}$ at the sub-saturation density of $\rho_c=0.1$ fm$^{-3}$. We note that the value of the surface-symmetry coefficient $a_{\rm ss} $ increases linearly with the value of $L_c$ in general, with which the corresponding value of $L_c$ for different macroscopic-microscopic and liquid-drop models could be estimated.

\begin{figure}
\includegraphics[angle=-0,width= 0.7 \textwidth]{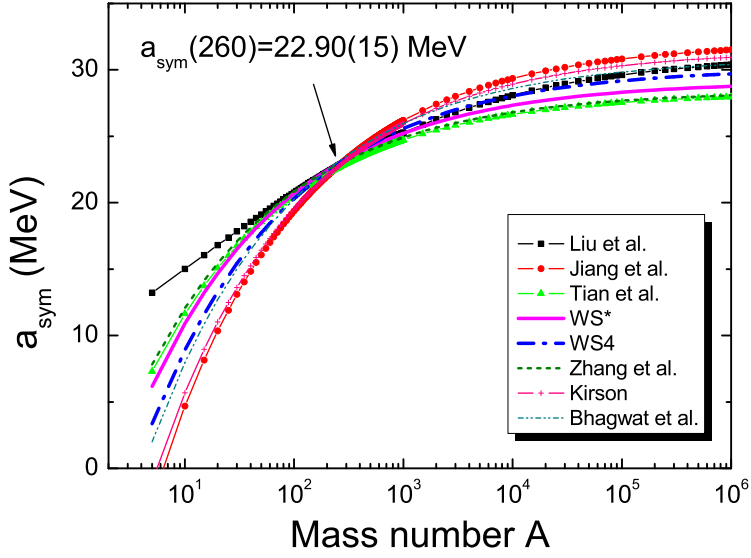}
 \caption{(Color online) Symmetry energy coefficients of nuclei extracted from nuclear masses.}
\end{figure}

To illustrate the importance of the slope $L_c$ and the density $\rho_c$, we simultaneously investigate the symmetry energy coefficients $a_{\rm sym}(A)$ extracted by using various liquid drop and macroscopic-microscopic mass models together with the measured masses of nuclei in this work.
It is known that only one-third of the nucleons in heavy nuclei occupy the saturation density area \cite{Khan12}. Consequently, nuclear
observables related to the average properties of nuclei, such as masses or radii, constrain the equation of
state not at the saturation density but rather around the so-called "critial" density $\rho_c\approx 0.1$ fm$^{-3}$. In Refs.\cite{Liu2010,Jiang12,Tian14}, the authors determine the coefficients of the symmetry energy term in the liquid drop model, assuming the form $a_{\rm sym}(A)=J-a_{\rm ss} A^{-1/3} $ or  $a_{\rm sym}(A)=J / (1+\kappa A^{-1/3})$ \cite{Danielewicz} and subtracting the Coulomb and Wigner terms from the measured binding energies of nuclei. The obtained symmetry energy coefficients $a_{\rm sym}(A)$ are shown in Fig. 5. Here, we also show the corresponding  $a_{\rm sym}$ adopted in five different macroscopic-microscopic mass models \cite{Wang14,Wang10a,Zhang14,Kir08,Bhag10} for comparison. Because the nuclear symmetry energy in nuclear masses is highly correlated with the Coulomb energy, the Wigner energy, and as well as the shell corrections, one can see from Fig. 5 that the extracted symmetry energy coefficients are quite different for light nuclei and the nuclei with huge numbers of nucleons. However, for nuclei with $A=260$, all these different models give quite similar predictions $a_{\rm sym}=22.90\pm 0.15 $ MeV. With the ETF2 approach, the corresponding result from SkM* is $a_{\rm sym}=22.95 $ MeV, which is in good agreement with the results in Fig. 5.

 \begin{figure}
\includegraphics[angle=-0,width= 0.7 \textwidth]{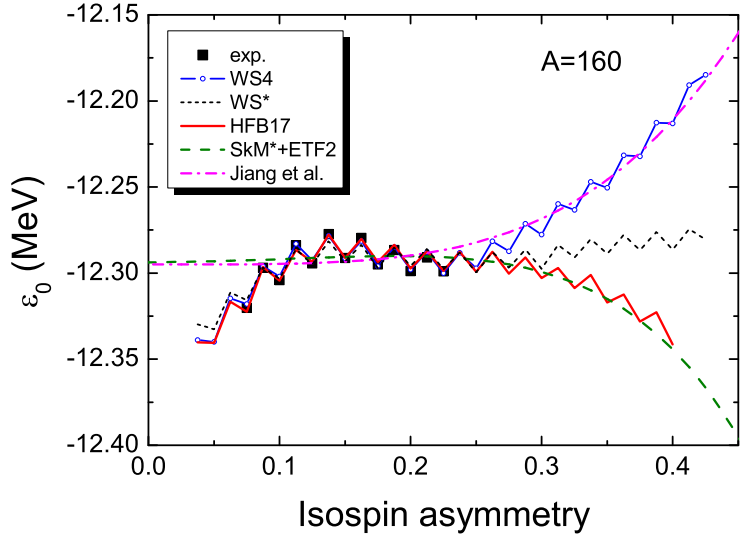}
 \caption{(Color online) Binding energy per particle after removing the Coulomb energy, Wigner energy and traditional symmetry energy.}
\end{figure}

As mentioned previously, the predicted masses of extremely neutron-rich nuclei are quite different from the WS4 and HFB17 models. To understand the large deviations, we analyze the influence of the $I^4$ term on the masses of neutron-rich nuclei. According to the well known liquid-drop formula, the ground state  energy of a nucleus is expressed as
\begin{eqnarray}
E_b \approx e_0 A +  E_C + E_W  + a_{\rm sym} I^2 A + a_{\rm sym}^{(4)} I^4 A +...,
\end{eqnarray}
with $e_0= a_v+a_s A^{-1/3}$ for the volume and surface terms, the Coulomb energy $E_C=0.71Z^2/A^{1/3}(1-0.76Z^{-2/3})$ and the Wigner energy $E_W=47|I|$ \cite{Sat97}. After removing the Coulomb energy, the Wigner energy, and the traditional symmetry energy ($a_{\rm sym} I^2 A$) from the total energy $E_b$ of a nucleus, one obtains the binding energy per particle of finite nuclei
\begin{eqnarray}
\varepsilon_0 \approx e_0  + a_{\rm sym}^{(4)} I^4,
\end{eqnarray}
neglecting the microscopic corrections in nuclei.
In Fig. 6, we show the calculated $\varepsilon_0$ for nuclei with $A=160$. The choice of $A=160$ is to avoid the influence of the shell effect as possible. The black squares denote the experimental data. The circles, short-dashed and solid curves denote the predictions of WS4 \cite{Wang14}, WS* \cite{Wang10} and HFB17 \cite{HFB17}, respectively. We take $a_{\rm sym}(160)=21.7$ MeV in the calculations according to the predictions of the eight models in Fig. 5. In addition, we show the results (dashed curve) of ETF2 together with SkM* for comparison. Here, the results of SkM* is shifted by 0.17 MeV to reproduce the experimental data. One can see that for nuclei with known masses, all the three mass models WS4, WS* and HFB17 can reproduce the experimental data remarkably well. However, the trend becomes quite different for extremely neutron-rich nuclei. The calculated $\varepsilon_0$ from the HFB17 model decreases with the isospin asymmetry, whereas the result of WS4 increases with the isospin asymmetry. The trend of $\varepsilon_0$ from the Skyrme force SkM* is similar to those from HFB17 which is based on the Skyrme force Bsk17 \cite{HFB17}. With the help of the ETF2 approach, we obtain the coefficient of the $I^4$ term $a_{\rm sym}^{(4)}=-2.84$ MeV for nuclei with $A=160$ by adopting SkM*. The corresponding values of the coefficient $a_{\rm sym}^{(4)}$ for nuclei with mass number $A=208$ and $A=40$ are also listed in Table I, denoted by $a_{\rm sym}^{(4)}(208)$ and $a_{\rm sym}^{(4)}(40)$, respectively. For almost all 36 selected Skyrme forces, the coefficients $a_{\rm sym}^{(4)}$  for finite nuclei are negative. In the WS4 model, the surface diffuseness correction is taken into account for unstable nuclei, which causes the enhancement of the symmetry energy for nuclei approaching drip lines and thus leads to the enhancement of $\varepsilon_0$ at large isospin asymmetry region. In the WS* model, neither the surface diffuseness correction nor the $I^4$ term is considered. Therefore, the results of $\varepsilon_0$ from WS* look flat in general. The decrease of the value of $\varepsilon_0 $ at small isospin asymmetry ($I < 0.1$) in the macroscopic-microscopic and HFB17 calculations could be due to the influence of shell effects around magic number $N=82$.

Very recently, Jiang et al. \cite{Jiang14} extracted the coefficient of the $I^4$ term with the double difference of the symmetry energy term together with the measured masses (AME2012) \cite{Audi12}, and a value of $a_{\rm sym}^{(4)} = 3.28$ MeV was obtained. Here, we also show in Fig. 6 the trend of $\varepsilon_0$ with $a_{\rm sym}^{(4)} = 3.28$ MeV (dot-dashed curve). One sees that the trend of $\varepsilon_0$ from Jiang et al. are close to those from the WS4 model with which the rms deviation to 2353 measured masses \cite{Audi12} is only 298 keV. With the same approach proposed in Ref. \cite{Jiang14}, we obtain $a_{\rm sym}^{(4)} = -3.07 $ MeV for the HFB17 mass model, which is generally consistent with the result of the ETF2 approach. For $^{168}$Sn ($I\simeq 0.4$) mentioned previously, the contribution of the $I^4$ term to the binding energy is $a_{\rm sym}^{(4)}I^4 A\approx \pm 13$ MeV according to the values of $a_{\rm sym}^{(4)}$ in HFB17 and WS4. It seems that the opposite values for the coefficient of $I^4$ term used in the HFB17 and WS4 models result in the large difference at the predictions of the masses of heavy nuclei approaching the neutron drip-line.

\begin{center}
\textbf{IV. SUMMARY}
\end{center}

Based on the extended Thomas-Fermi (ETF2) approximation together with the restricted density variational method, the symmetry energy coefficients of finite nuclei including the coefficients of the $I^4$ term have been systematically investigated with 36 different Skyrme energy density functionals. From nuclei with mass number $A=20$ to the nuclei with huge numbers of nucleons, of the order of $10^6$, we study the mass dependence of the symmetry energy coefficients. With the extrapolations from the calculated results for finite nuclei, the asymptotic values of the symmetry energy coefficient $J$,  of the binding energy per particle $e_0 (\infty)$ and of the coefficient $a_{\rm sym}^{(4)}(\infty)$ in the $I^4$ term when $A\to \infty$ have been compared with the corresponding values from the asymmetric nuclear matter. The obtained results from the finite nuclei are very close to those from the nuclear matter, which indicates that the ETF2 approach is reliable for extracting the symmetry energy coefficients of finite nuclei, because the Coulomb energies, the Wigner energies and the shell corrections can be "cleanly" removed in the calculations.

With the help of the ETF2 approach, the reference density of finite nuclei is also investigated. We find that the extracted symmetry energy coefficients $a_{\rm sym}(A)$ from different liquid-drop and macroscopic-microscopic models for heavy nuclei such as nuclei with $A=260$ are in good agreement with each other, $a_{\rm sym}(260)=22.90\pm 0.15 $ MeV, although the uncertainty for very light nuclei and those for nuclear matter is relatively large. The ETF2 approach with SkM* gives a similar result $a_{\rm sym}(260)=22.95 $ MeV and the corresponding reference density is $\rho_A\simeq 0.1$ fm$^{-3}$.

In addition, we analyze the large deviation between the HFB17 and WS4 model for the predictions of the masses of extremely neutron-rich nuclei. From the predicted binding energy per particle for nuclei with $A=160$ after removing the Coulomb energy, Wigner energy and traditional symmetry energy, we find that the HFB17 model gives a negative value for the coefficient $a_{\rm sym}^{(4)}$ of the $I^4$ term, whereas the WS4 model gives a positive value due to considering the surface diffuseness correction of nuclei. The extracted result from nuclear masses with the double difference to the symmetry energy term suggests a positive value for $a_{\rm sym}^{(4)}$. The negative values for $a_{\rm sym}^{(4)}$ from the systematic study of all 36 different Skyrme forces imply that the experimental data used in the different fitting protocols do not constrain this part of the standard Skyrme functionals and/or new terms should be included in future Skyrme functionals.

\begin{center}
\textbf{ACKNOWLEDGEMENTS}
\end{center}

This work was supported by National Natural Science Foundation of
China (Nos 11275052, 11365005, 11422548, 11475004, 11225524, 11305101) and the 973 Program of China (Grant Nos. 2013CB834401). N. W. is grateful to Lie-Wen Chen and Li Ou for valuable discussions, and acknowledges the support of the Open Project Program of State Key Laboratory of Theoretical Physics, Institute
of Theoretical Physics, Chinese Academy of Sciences, China (No. Y4KF041CJ1).

\end{document}